\documentclass{he_symp}
\usepackage{psfig,graphicx,epsfig}
\usepackage{color}
\usepackage{amsmath,amssymb,epic,eepic,array}
\begin{document}
\renewcommand{\FirstPageOfPaper }{ 221}\renewcommand{\LastPageOfPaper }{ 229}

\title{A current circuit model of pulsar radio emission}
\author{Th Kunzl\inst{1,2}, H. Lesch\inst{2} \and A. Jessner\inst{3}}  
\institute{Max--Planck--Institut f\"ur Quantenoptik, Hans-Kopfermann-
Stra{\ss}e, 85748 Garching, Germany \and Institut f\"ur Astronomie und
Astrophysik der Universit\"at M\"unchen, Scheinerstr.~1, D-81679 M\"unchen
\and Max--Planck--Institut f\"ur Radioastronomie, Auf dem H\"ugel 69, D-53121 Bonn}

\authorrunning{Kunzl et al.}
\titlerunning{A current circuit model of pulsars}

\maketitle

\begin{abstract}

We present the outline of a new model for the coherent radio emission
of pulsars that succeeds in reproducing the energetics and brightness
temperatures of the observed radio emission from the observationally deduced
distances of 50-100 pulsar radii above the neutron star in a narrow region.

The restrictions imposed by energy conservation, plasma dynamics of
the coherent radiation process and propagation effects are used to
apply the action of a plasma process like coherent inverse Compton scattering
(CICS) (see Benford, 1992). In accordance with our findings (Kunzl et al.~1998a)
 this process requires Lorentz factors of about 10 which are lower than
in most other radio emission models. This implies that no significant pair
production can take place near the surface and we expect charge densities 
close to the Gold\-reich-Julian value (Goldreich \& Julian (1969)).

To fulfill the energetic and electrodynamic constraints 
the model requires constant
re-acceleration in dissipation regions which can be interpreted as a voltage drop
similar to that in a resistive current circuit built of a battery, connecting
copper wires and a resistive load.

Using the emission heights for PSR 0329+54 published by Mitra and Rankin (2002) 
and the spectral data from Malofeev et al.~(1994) we find that a constant depth
 of the dissipation region  of  about $2 r_{\rm NS}$ can account for the observed 
luminosities and spectral behavior.

The extremely high peak fluxes in the substructures of single pulses can be explained by
beaming effects as discussed in Kunzl et al.~(1998b).

\end{abstract}

\section{Introduction}

Pulsars are generally accepted to be rotating neutron stars having
$10^4-10^8$T magnetic fields that induce electric potential drops
along the magnetic field lines which allow efficient particle
acceleration of thermally emitted electrons in the region of open
field lines (Gold\-reich and Julian (1969); Sturrock (1971); Ruderman and
Sutherland (1975), Arons (1981)).
Although the emitted radio luminosity in comparison to the spin-down power
of a rotating neutron star is tiny (only the $10^{-5}$- $10^{-6}$ part of
the spin-down power is emitted in the radio regime)
one of the most striking features about pulsars is their very intense
pulsed radio emission, the physical origin of which is still poorly
understood. The wealth of observational data
(intensity, polarization, shape of the radio
spectrum) clearly indicates that the pulsar radio emission cannot be of
incoherent origin, but must be due to some coherent radiation mechanism (e.g. Lyne and Graham-Smith 1998;
Kramer, Wex and Wielebinski (2000) and references
therein). Especially at low frequencies (below 1 GHz) a very high degree of
coherence is needed to explain the observed intensities in terms of brightness temperatures.
Therefore a pulsar emission model has to explain how a large number of phase
coupled particles can be stimulated to radiate in a coherent manner.

Currently the most favored models for radio emission mechanism are curvature
radiation (e.g. Rankin (1992); Radhakrishnan and Rankin (1990); Gil (1992)) or
maser process due to strong turbulence, like the free electron maser,
(hereafter FEM) (e.g. Melrose (1978), Rowe (1992a,b)).

Since there are severe difficulties with coherent curvature radiation (Lesch et 
al.~(1998) pointed out that the energetic requirements can hardly be
matched, Melrose (1992) discussed the absence of an appropriate bunching
mechanism to build up a coherent curvature radiation) we propose a process directly coupled to the
most significant eigenfrequency of the magnetospheric plasma: the electron plasma frequency.
It has been shown by Melrose (1978) that such an amplified linear
acceleration emission (ALAE) process could produce sufficient growth rates
i.e.~is fast enough to explain the observations. But it needs low Lorentz
factors of the emitting particles to work. This model requires low energy particles
with energies of a few MeV.

Recently, support for such a low energy scenario came from various criticisms on
the so-called standard model with an inner gap where a highly relativistic pair
plasma is created:

It was argued by Kunzl et al.~(1998a) and confirmed by Melrose and Gedalin (1999, 2000)
that the observed radio emission would not escape from a magnetosphere
filled with a dense relativistic pair plasma. Additional problems with the
production of high energy particles close to the pulsar were discussed by
Arendt \& Eilek (2000), Lesch et al.~(2000) and Jessner et al.~(2001).  They commonly
concluded that the assumption of strong electric fields at the pulsar surface that
ultimately result in ultrarelativistic particle energies is at least questionable.

Thus, we introduce an entirely different approach here, which
has first been promoted by Shibata (1991) for an analysis of the global
magnetospheric structure of rotation-powered neutron stars.
In contrast to the standard model assuming a shielding of
the electric fields by massive pair production we propose that in the inner
region of the magnetosphere (i.e.~the open field lines inward of the neutral
surface) the situation is similar to the multiple scatterings in a solid state conductor.
In other words we have a
current whose time average is the same at all cross sections through the open
field line region. There is neither acceleration close to the
surface (as indicated by our recent findings (Jessner et al.~2001) nor
deceleration in the dissipation region (as indicated by the very small
dissipation rate given by small the radio luminosity in comparison to the total
spin down power).

In this conducting wire model any dissipation region corresponds to a resistance
where electric potentials are present. The "deceleration" caused by the losses is
balanced by the "acceleration" due to the electric field. The most important
consequence of this scenario is that the energy loss per particle is no longer
limited by the average kinetic energy of the electron in the same way as the kinetic
energy of the electrons in a copper wire current circuit is no indicator
for the dissipated energy. The one-dimensional approach with a constant voltage
applies, because the resistivity perpendicular to the magnetic field is small
inside the pulsar (to allow sufficient supply of charged particles for a
Goldreich-Julian DC current) and large enough in the magnetosphere (to avoid
"leakage", meaning current closure across the field lines in the inner
magnetosphere) (see Ewart et al.~1975; Itoh 1975).

Our aim is now to obtain the constraints on the dissipation field in such a
model and see whether we can explain the luminosities in the radio frequency
range at the emission heights given by the observations from a thin radiating
layer.

\section{The model}

{In a current-carrying collisionless plasma waves are known to introduce
resistivity via fluctuating electric fields. These fields present the deviations
from the ideal plasma state with an infinitely high
electrical conductivity. For an incoming current $I$ the locations of a
significant high plasma wave intensity act as a resistance $R$.
The emitted power of such a resistive region is given by $L=I^2 R$.} In a time
averaged picture the current has to be divergence-free. So if a DC current flows
through a region in which plasma waves generate a resistance the particles lose their kinetic energy meaning
that their velocities decline. Although for relativistic particles this effect
is small it cannot be entirely neglected.

In a dense pair plasma this energy loss of the particles presents no
difficulties since the velocities of the two components can adjust in an
adequate way to fulfill both energy and current conservation. However, in a
single charge plasma deceleration has to be accompanied by a density rise. So if
the particle beam would gradually lose energy the relative density would have to
increase all the time. This is totally incompatible with the outer boundary
condition of a Goldreich-Julian magnetosphere.

The latter point is the main difference to the argumentation by Shibata (1997)
who also assumes a current outflow determined by the system. In Shibata's model
curvature effects causes space charges to accumulate that cannot be removed by acceleration on
some field lines as the particle velocity is limited by $c$. We assume that
these curvature effects do not play a significant role as they will be balanced
by particles of the opposite charge drifting inwards from the neutral line and
therefore only deceleration by losses will result in a space charge. This would
mean that radiation losses produce a global deviation from the GJ case
if there was no acceleration of the particles. So such a description would
immediately turn out to be inconsistent (non-idealities induce electric fields
which accelerates or decelerates particles).

Nevertheless there is a consistent description of a single charge plasma
distribution  if we recall that the rotating neutron star induces the vacuum
electric field. The electric charges {\em on the neutron star surface} are
displaced in such a way that the field that has to be shielded by the GJ charge
and the field is actually anchored in the neutron star. Thus, the vacuum field
can be interpreted as a stationary distribution of infinitely heavy
positive particles whose strength is given by the entire pulsar system.

This means that electrons that suffer losses will
immediately be dragged on by the electric field that builds up in such a way
that just an electrostatic wave is produced. A particle transversing this wave
is accelerated by this non-vanishing potential drop and looses the gained energy
by non-thermal radiation.

In this case the particles have to pile up which means that
their velocity has to
decrease. Where the density goes down, the velocity has to increase accordingly
to keep the current stable (${\rm div}\,\vec{j}=0$).

Then the resulting electric field is able to accelerate particles and varies
with half the wavelength of the density variation (as follows directly from the
Poisson equation).

Such a system is similar to a battery whose poles are the region of the
open field lines (acting as the negative pole in the parallel rotator) and the
edge of the polar cap being the anode. If a particle starts close to the pole
and returns on the last closed field line it encounters a voltage drop of
typically $10^{13}$ V. But, of course, this does not mean that the particle has
to be accelerated {\bf by the full potential in one step}  anywhere along its
path.

\begin{figure}[htbp]
 \centerline{\psfig{file=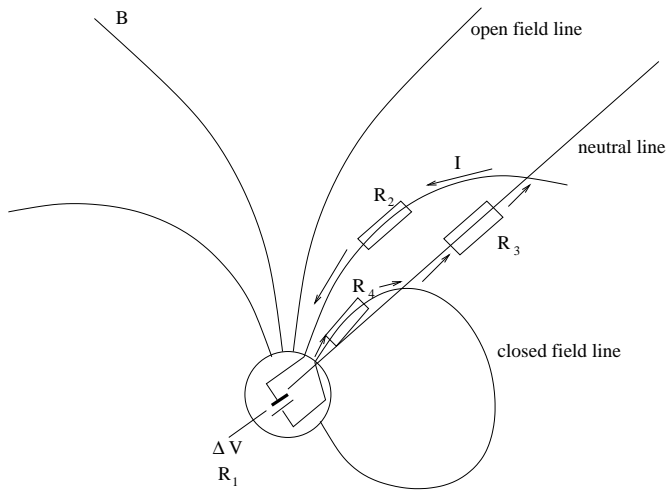,width=8.8cm,clip=}}
 \caption{\label{SB} Schematic view of the current circuit model. The current is
 determined by the applied voltage $\Delta V$ and the resistances $R_1$ to $R_4$. $R_1$ is
 the battery's resistance which is assumed to be small. $R_2$ stands for plasma
 resistance effects, mainly the losses by radio emission. The vastly dominating
 resistance is $R_3$ which limits the current to a relativistic GJ current over
 the two polar caps. In $R_4$ we combine other possible resistances which are
 also assumed to be small (e.g. curvature radiation losses or similar effects)}
\end{figure}

As will be motivated in sect.~7 it is likely that the outer magnetosphere
itself provides a constant minimum resistance which clearly dominates all plasma induced
resistances so that the latter are only minor corrections. In other words, if
the main voltage drop occurs close to the light cylinder almost all of the
energy is dissipated there and transformed into high energetic radiation and
ultrarelativistic particles. A schematic view of the model is shown in fig.~\ref{SB}..

Of course, there are two basic assumptions in this model, which have to be
checked carefully. On the one hand, the pulsar is described as a voltage
source. This description requires that the generator's internal resistance
is negligible compared to the total resistance. It will be shown in section 6
that the maximum current allowed would be a highly relativistic Goldreich-Julian
current over the entire two polar caps. The corresponding resistance is about
$5\,\Omega$ which is huge in comparison with the transverse resistance of about
$10^{-8}\,\Omega$ (Itoh, 1975)
Thus the first assumption appears to be quite reasonable.

The second assumption, namely the one-dimensional current flow in the inner
magnetosphere is also very natural. The conductivity can be expressed by
a mean collision time $\tau$ using the formula

\begin{equation}
 \sigma=\frac{ne^2\tau}{m_{\rm e}\varepsilon_0}\,.
\end{equation}

The transverse collision time $\tau$ can be estimated by $1/\omega_{\rm c}$
(the inverse gyro frequency) whereas the collision time for the parallel current
cannot exceed the inverse plasma frequency $1/\omega_{\rm p}$ (see below). As in the inner
magnetosphere $\omega_{\rm p}\ll\omega_{\rm c}$, the one-dimensional approach is
appropriate. Another way of confirming this is to calculate the maximum
(transverse) drift velocity in the inner magnetosphere. Taking the maximum
plasma wave electric field (see below) and typical pulsar parameters ($P=0.5$ s,
$B_0=10^8$ T) we find a drift velocity of

\begin{equation}
v_{\rm drift}=0.359 \frac{\rm m}{\rm s}\,x_{\rm em}^{3/2}\gamma^{1/2}
\end{equation}

which is far below $c$ for typical emission heights and particle energies used
in the model ($x_{\rm em}\approx 50$, $\gamma=10$). Particles move almost
perfectly along the field lines as long as relativistic parallel outflow is
possible (on the neutral line the situation is a bit different as a pure
electron flow cannot cross this border easily).

An important consequence of our description is that the total energy a particle
radiates along its path is in no way limited by its own kinetic energy as the
actually dissipated energy comes from the inductor, {\em i.~e.}~the battery.
Here, the analogy to a battery with a bulb attached through copper wires can be helpful. In
a metal conductor the current carrying electrons have velocities of only some
millimeters per second which means a kinetic energy of only $10^{-16}$eV
 because of the small path length of electrons in a solid state conductor.
 Nevertheless each of them converts a total of 1.5 eV into heat on its way from the cathode
to the anode. The only restriction which is given by the kinetic energy is the
potential difference of the electrostatic field in one elementary radiation
cell. As we will point out in sect.~4 this limit directly leads to the maximum
possible resistivity in the plasma and therefore is never exceeded anyway.

The accelerating fields occurring in the outer magnetosphere do not
invalidate the current circuit model, however as the circuit is open (high
resistance) near the light cylinder.

Another advantage of this description is that our model uses particle inertia as
the source for an accelerating field near the pulsar surface to get relativistic
particles. So a very small acceleration region has to occur as the electrons
have non-relativistic velocities inside the neutron star; therefore an overdense
thin charge layer develops just above the surface where the electrons reach
their terminal energy.

Therefore is no need for a "gap" (meaning a starved region near the neutron
star) but inertia naturally provides the fields necessary to reach the mildly
relativistic energies, and the terminal energy is determined by the global
current instead of some large-scale space charge field (cf.~sect.~6).

\section{Application to radio emission}

In the following we derive a quantitative model for the radio emission of
pulsars.
The particular aim is to explain the energetics of the radiation, from observed
or observationally derived pulsar parameters such as

\begin{itemize}
 \item{the period $P$}
 \item{the dipolar surface magnetic field $B$}
 \item{the brightness temperature $T_{\rm B}$ of the low frequency emission}
 \item{the emission height $x_{\rm em}$}
 \item{the total luminosity in the radio band $L$}
\end{itemize}

With these input parameters we derive the radial width of the emission region,
which should  be small to be consistent with observations. A
narrow radiation zone is suggested by both the highly structured emission in
space and time as seen by many microstructure experiments (Boriakoff, 1992).


\subsection{Plasma beams and coherent emission}

Emission processes in relativistic plasmas are always connected with the
electron plasma frequency, which is (Kunzl et al.~1998a; 
Melrose and Gedalin 1999)

$$\omega_{\rm pe}=\sqrt{{n e^2\over{\gamma\varepsilon_0 m_e}}}$$

As a very promising candidate for the origin of coherent radio
emission of pulsars we consider the collective inverse
Compton mechanism, which was detected first in
a number of laboratory experiments in which relativistic electron
beams with a number density $n_{\rm b}$ penetrate into a background plasma
with density $n_{\rm p}$ and in which extremely intense coherent microwave emission
has been observed (Kato et al.~1983; Levron et al.~1987; Yoshikawa et al.~1993;
1994). The detected radiation has been explained in terms of collective inverse
Compton scattering of strong Langmuir turbulence (Benford and Weatherhall 1992;
Benford 1992). Relativistic
electron beams interact with self-excited strong density fluctuations in the background plasma
if the beam-plasma-density ratio $n_{\rm b}/n_{\rm p}$ exceeds 0.1 for a beam Lorentz factor
of about 3. The beam electrons are bunched by the electric field of the density
fluctuations and radiate coherently in their reference system at
 $\gamma\omega_{\rm pe}$, transformed into the observer's
frame this gives an emitted frequency of about $\gamma^2\omega_{\rm pe}$.
The efficiency of this energy transfer of beam energy into photon energy is
up to 30\%.

The general scaling law from laboratory experiments for the relation between beam
Lorentz factor and density ratio is (Benford 1992)

$$\gamma_{\rm crit} = \left (1+{9\over{20}}{\rm lg}\left
({n_{\rm b}\over{n_{\rm p}}}\right)\right)^{-1}$$

\begin{figure}[htbp]
 \centerline{\psfig{file=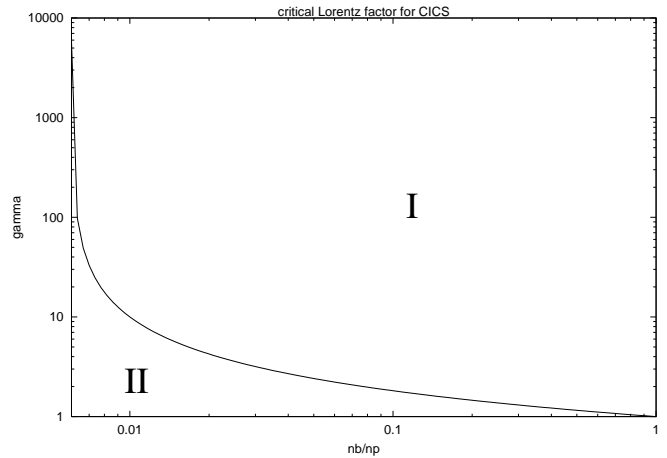,width=8.8cm,clip=}} 
 \caption{\label{CICS} Minimum Lorentz factor that allows coherent inverse
 Compton scattering (CICS) as a function of the ratio of beam density $n_{\rm
 b}$ to  background plasma density $n_{\rm pl}$ (see Benford, 1992). For strong
 beams ($n_{\rm b}\approx n_{\rm pl}$) even low relativistic particles can cause
 coherent radiation (region I). In region II no coherent emission is possible.}
\end{figure}

As shown in Fig. \ref{CICS} for large beam energies the required density ratios
is quite small, whereas for small beam energies the beam density must be
comparable to that of the background plasma to drive collective inverse Compton
scattering. For a given beam energy one can calculate the required density
ratio.  

The underlying physics of this mechanism has been investigated in all
details and the general consensus is, that a relativistic electron
beam with sufficient energy and particle density penetrates into
 a background plasma the collective inverse Compton scattering is the
unavoidable coherent radiative power output (Benford 1992 and references therein). Based on
these findings we transfer this process into a neutron star magnetosphere
and investigate, whether this process matches the required properties.

Collective inverse Compton scattering is a nonlinear scattering
processes whose radiative output in first order does not depend on the
origin of the density fluctuations. In contrast, direct excitation mechanisms
like maser or beam radiation depend sensitively on the form of the
beam energy distribution and the time evolution of the beam
(Ursov \& Usov 1988; Melrose \& Gedalin 1999; Melrose et al.~2001).

For comparison with alternative processes, we give a general scaling of the emitted frequency with the
nonrelativistic plasma frequency $\omega_{\rm pe}^{(0)}$ for the different
radiation mechanisms

\begin{equation}
 \nu=\frac{\omega_{\rm pe}^{(0)}}{2\pi}\gamma^{\alpha/2}\,.
\end{equation}

In this notation the value of $\alpha$ for the collective inverse Compton scattering
is 3, whereas for direct emission it is 1. In the next chapter we will
debate an argument why the description of the pulsar system as a resistive
current circuit makes $\alpha=1$ less likely than a higher value. Inserting
the expressions for the Goldreich-Julian density and the plasma frequency, we
can solve this equation for $\gamma$:

\begin{equation}
 \label{gamomega}
 \gamma=\left[\frac{\omega_{\rm p}(x_{\rm em})}{2\pi\nu}\right]^{-2/\alpha}=
 \left(\frac{B_0\Omega e}{2\pi^2\nu^2m_{\rm e}x_{\rm em}^3}\right)^{-1/\alpha}\,.
\end{equation}

According to the usual estimates, an incoherent process from an electron with a
Lorentz factor of $\gamma$ cannot produce radiation with a brightness
temperature exceeding $10^{10}{\rm K}\gamma$. However, in the case discussed
here, this argument is not applicable, because there is no absorption. The
emitted power of a radiation process is not limited any more if the radiation
produced further inside is not absorbed on its way out of the magnetosphere.

Consider a cylindrical plasma column of radius $R$, length $l$, particle density
$n$ and single particle power output $P_{\rm single}$ in forward direction. Then
the total power adds up to

\begin{equation}
 P_{\rm ges}=l R^2\pi n P_{\rm single}.
\end{equation}

Thus the brightness temperature can reach arbitrarily high values on the
condition that $l$ is large enough no matter, whether the emission process is
coherent or not.

Consequently we use a different estimate for the brightness temperature. The
average flux is received from an area that corresponds to the polar cap moved and
expanded along the field lines up to the emission height:

\begin{equation}
 \label{Acap}
 A=A_{\rm cap}x_{\rm em}^3\,.
\end{equation}

The total radio luminosity over a frequency band of $\Delta\nu$ around the
frequency $\nu$ can therefore be equated to the thermal flux of a blackbody with
the appropriate brightness temperature. For broadband emission ($\nu=\Delta\nu$)
and using the definition of the brightness temperature we find

\begin{equation}
 L=F_\nu\Delta \nu A_{\rm cap}x^3 \approx \frac{2\pi\nu^3}{c^2}k_{\rm B}T_{\rm
 B} A_{\rm cap}x_{\rm em}^3
\end{equation}

which yields a brightness temperature of

\begin{eqnarray}
 \label{Helltemp}
 &T_{\rm B}& = \frac{Lc^2}{2\pi\nu^3 k_{\rm B}A_{\rm cap}x_{\rm em}^3}= \\
 &=& 1.87\cdot 10^{23}\, {\rm K} \left(\frac{L}{10^{20}\,{\rm W}}\right)
 \left(\frac{\nu}{400\,{\rm MHz}}\right)^{-3}\left(\frac{P}{\rm s}\right)
 \left(\frac{x_{\rm em}}{50}\right)^{-3}\,. \nonumber
\end{eqnarray}

Note that there is no additional geometric factor as the estimate of the total
radio luminosity already averages over some typical pulse profile.
The values for the brightness temperature found in (\ref{Helltemp}) are consistent
with observationally deduced results (Sutherland, 1979, Kramer, 1995) as the
same principal method has been used in both cases.

\smallskip

High coherence is required for a different reason, namely that the radiating
volume is small enough. In the next step we show that our model is able to
reproduce these fluxes under the assumption of a thin layer emitting
highly coherent radiation, but where the coherence factor is kept below the
limit imposed by the model. As already mentioned, we propose a mechanism of the
relativistic plasma emission type.

For the discussed process, particles are scattered by strong, nonlinear plasma
waves (solitons) excited by the two-stream instability. Such a radiation
mechanism can be described as an inverse Compton scattering (ICS) of solitons by
relativistic electrons.

For each particle coherent ICS radiates a power of

\begin{equation}
 \label{Prad1}
 P_{\rm rad}^{\rm coh}=N\sigma_{\rm T}c\cdot \frac{E_{\rm
 wave}^2\varepsilon_0}{2} \gamma^2
\end{equation}

with the wave electric field $E_{\rm wave}$ if $\hbar\omega_{\rm wave}\ll
m_{\rm e}c^2$, which is the case for radio emission\footnote{For
higher wave energies the cross section is no longer described by $\sigma_{\rm
T}$, so that the Klein-Nishina cross section ($\sigma_{\rm KN}\approx
\sigma_{\rm T}({\rm {ln}}\,\gamma) / \gamma $) must be used.}. $N$ is the
coherence number, and $\gamma^2$ comes from the Lorentz transfomation of $E_{\rm
wave}$. As long as the wave electric field is small, the wave can be treated
linearly as a small density fluctuation. But in an unstable situation the wave
grows exponentially. The growth saturates quickly because the wave will also
accelerate particles as a back reaction. Thus, the wave growth has to
be calculated using non-linear methods (Benford \& Weatherall, 1992). As a
result, the strongest possible wave has an electrostatic field energy comparable
to the kinetic energy of the plasma. In the extreme, these two energy densities
are equal

\begin{equation}
 n_{\rm GJ}\gamma m_{\rm e}c^2=\frac{1}{2}\varepsilon_0 E_{\rm wave}^2
 \label{Maxfeld}
\end{equation}

which means (\ref{Prad1}) can
be rewritten as

\begin{equation}
 \label{Prad2}
 P_{\rm rad}^{\rm coh}=N\sigma_{\rm T}cn\gamma^3 m_{\rm e}c^2\,.
\end{equation}

As particles are not assumed to change their longitudinal momentum in the radio emission region,
we can interpret the radiated power as an electric dissipation field (which is
the field necessary to balance radiation losses). Its strength is

\begin{equation}
 \label{Ediss}
  E_{\rm diss}=\frac{P_{\rm rad}^{\rm coh}}{ec}=\frac{2N\sigma_{\rm T}\Omega B_0
  \varepsilon_0 m_{\rm e}c^2}{e^2x_{\rm em}^3}\gamma^3\,.
\end{equation}

\subsection{Extension of the radiation region and the maximum luminosity}

Now we calculate the radial width of the radiation zone. This is an important
test of the model, as the very small RFM taken from the observations requires
a fairly narrow emission zone. Apart from that, the highly modulated radiation
observed in the radio band would be hard to explain for large widths.

In (\ref{Ediss}) the only the free parameter left is $N$. However, this number
cannot directly be fixed from the brightness temperature as mentioned
before. Nevertheless there are upper limits for the coherence factor which serve
to derive a minimum radial width of the emission region.

As we consider a plasma wave being the cause of coherence, the maximum volume of
coherently radiating particles is a sphere of around one Debye volume. Taking
into account relativistic effects, Melrose (1992) finds a maximum coherence
volume of

\begin{equation}
 V_{\rm coh}=\left(\frac{c}{\omega_{\rm p}}\right)^3\frac{\gamma^2}{\pi}
\end{equation}

and therefore

\begin{eqnarray}
 \label{Nbunch}
 N_{\rm max} &=& n_{\rm GJ}\left(\frac{c}{\omega_{\rm
 p}}\right)^3\frac{\gamma^2}{\pi}=\sqrt{\frac{m_{\rm e}^3
 c^6\varepsilon_0\gamma^7x_{\rm em}^3}  {2\Omega}B_0e^5}= \nonumber \\
 &=& 6.41\cdot 10^{13}\gamma^{7/2}\left(\frac{P}{\rm s}\right)^{1/2}
 B_8^{-1/2}\left(\frac{x_{\rm em}}{50}\right)^{3/2}\,.
\end{eqnarray}

Another upper limit is found by equating the particle and the field
energy density, since in the optimal case those quantities are comparable. More
detailed calculations (Lesch \& Schlickeiser, 1987) show that the field energy
density is less than half of the particle energy density (as already used in
eq.~(\ref{Maxfeld})).

Inserting this result into (\ref{Ediss}) we find a maximum coherence number of

\begin{eqnarray}
 \label{Nplasma}
 N &=& \sqrt{\frac{e^3x_{\rm em}^3}{2\gamma^5\Omega B_0 m_{\rm
 e}c^2\varepsilon_0^2\sigma_{\rm T}^2}}=\nonumber \\
 &=& 3.79\cdot 10^{15}\,\gamma^{-5/2}\left(\frac{P}{\rm s}\right)^{1/2}
 B_8^{-1/2} \left(\frac{x_{\rm em}}{50}\right)^{3/2}\,.
\end{eqnarray}

Comparing the maximum coherence numbers from (\ref{Nbunch}) and (\ref{Nplasma})
we find that the latter is smaller whenever $\gamma\geq 1.97$. Therefore it is a
good approximation to use (\ref{Nplasma}) for determining the maximum coherence
number.

For estimates of the radial extension of the radio emission region we take
the total radio power and calculate the radiating volume necessary to reproduce
the observed luminosities.

The total energy loss in a resistive current reads

\begin{equation}
 \label{Lum}
 L=\int\limits_{V_{\rm rad}} E_{\rm diss}j\,{\rm d}V
\end{equation}

where $V_{\rm rad}$ denotes the dissipation volume.

$E_{\rm diss}$ is given by (\ref{Ediss}) and (\ref{Nplasma}), $j(x)=n_{\rm
GJ}(x)ec$, and by assuming that the non-ideality extends over the entire
cross-section of open field lines; using a dipolar geometry and neglecting
angular corrections to the GJ- density all quantities in (\ref{Lum}) only depend
on the radial coordinate. The volume element d$V$ can be transformed to

\begin{equation}
 {\rm d}V = r_{\rm NS}A_{\rm cap}x^3\,{\rm d}x\,.
\end{equation}

Here we have used (\ref{Acap}), as the cross-section is a simple function of
$x$. Therefore (\ref{Lum}) reduces to a one-dimensional integral.

Inserting all the previous expressions the integral reads

\begin{eqnarray}
 L &=& \left(\frac{2\pi^2\nu^2m_{\rm e}}{B_0\Omega e}\right)^{1/(2\alpha)}
 \sqrt{\frac{2\Omega B_0 m_{\rm e}c^2}{e}}\,\frac{2\Omega B_0 \varepsilon_0}{e}
 ec \nonumber \\
 & & r_{\rm NS}\,\frac{\pi \Omega r_{\rm NS}^3}{c}\,
 \int\limits_{x_{\rm em}}^{x_{\rm em}+\tilde{l}_0}x^{3/(2\alpha)}x^{-3/2}\,
 {\rm d}x
\end{eqnarray}

where $\tilde{l}_0$ denotes the radial extension of the radiation region in
units of $r_{\rm NS}$, whereas $x_{\rm em}$ stands for the height of its inner
edge in pulsar radii.

Evaluating the integral and inserting numbers for a frequency of $\nu=400$ MHz
we find

\begin{equation}
L=5.56\cdot 10^{22}{\rm W}\left(\frac{P}{\rm s}\right)^{1/\alpha-5/2}
B_8^{3/2-1/\alpha} F_\alpha
\end{equation}

\noindent
where

\begin{equation}
 F_\alpha=0.0286^{1/\alpha} \left\{ \begin{array}{cc} \tilde{l}_0 & {\rm
 for}\;\alpha =1 \\ 4 \left[\left(x_{\rm em}+\tilde{l}_0\right)^{1/4}-x_{\rm
 em}^{1/4}\right] & {\rm for}\;\alpha =2 \\
 {\rm ln}\frac{x_{\rm em}+\tilde{l}_0}{x_{\rm em}} &  {\rm for}\;\alpha =3
 \end{array}\right.
\end{equation}

\begin{figure}[ht]
 \centerline{\psfig{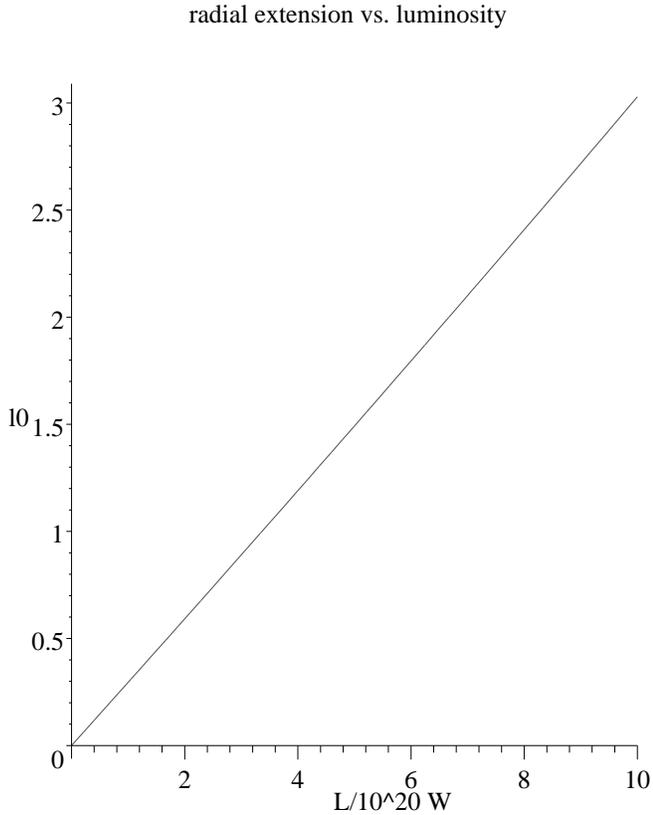}} 
 \caption{\label{l0vsL} Radial extension of the radio dissipation region (in
 units of pulsar radii) versus luminosity for a standard pulsar ($P=1$ s,
 $B_8=1$). The emission height  $x_{\rm em}$ has been set to 50 pulsar radii.}
\end{figure}

Solving this equation for $\tilde{l_0}$ in the case $\alpha=3$ renders

\begin{eqnarray}
 && \tilde{l_0}  = x_{\rm em}\\
 && \left\{ {\rm exp}\left[5.88\cdot 10^{-3}
 \left(\frac{P}{s}\right)^{13/6} B_8^{7/6} \left(\frac{L}{10^{20}\,{\rm W}}
 \right)\right]-1\right\} \nonumber
\end{eqnarray}

which, for typical parameters ($L=10^{20}$ W, $P=1\,{\rm s}$, $B_8=1$, $x_{\rm
em}=50$) yields $\tilde{l_0}=0.295$ which justifies the assumption of a narrow
emission region. The dependence of $\tilde{l_0}$ and $L$ is shown in
fig.~\ref{l0vsL}.

From the result above one can obtain sufficient luminosity even for slow
pulsars. However, for small $\Omega$ the radial extension of the emission region
increases. Especially, the corresponding electric field (\ref{Ediss}) can even
exceed the vacuum field ($E_{\rm vac}=\Omega B_0r_{\rm NS}x^{-4}$) locally.
In a shielded magnetosphere the kinetic energy density and by equipartition the dissipated electric field,
 may exceed the energy density of the vacuum field.

\section{Application to PSR 0329+54}

The pulsar PSR 0329 is an extensively researched object and one of the few where
the spatial orientation of magnetic and spin axis is well determined (Rankin, 1983).
Assuming that the profile peaks occur on the same field
lines for all frequencies enabled Mitra and Rankin (2002) to obtain revised
effective opening angles and  emission heights for PSR 0329+54
amongst others. 
A good fit to the surface opening angle (including relativistic corrections) 
for the field lines that provide the two outer peaks of 
the profile is $\phi_0=0.66^{\rm o}$. For any angle $\phi$ between those peaks we can
find an emission height of a profile at a specific frequency using the parametric 
equation of dipole field lines:
\begin{equation}
r(\phi)={r_{ns}\over f_1}{\sin(\phi)\over \sin(\phi_0)}
\end{equation}
Mitra and Rankin (2002) found that the emission heights range from about
500 km for frequencies above 1~GHz up to 1100 km for the
lowest frequency of 100~MHz.
Together with the known radio fluxes (Kramer et al.~1997) we were able to
match the described model to the data and determine the dissipation length
scales for frequencies between 100 MHz and 30 GHz. To enhance the realism of the
fit, the relativistic (Muslimov \& Tsygan, 1992) corrections to the
Goldreich-Julian densities and effective areas were also included. 
In this case we calculated the local Goldreich-Julian density through
\begin{equation}
n(r)=2 \epsilon_0 B_0 \left({r_{\rm NS} \over r }\right)^3 \cdot {f(r) \over f_1
\cdot \sqrt{1-{r_{\rm g}\over r}} }
\end{equation}
Here 

$$f(r)= -3({r \over r_{\rm g}})^3\cdot \left({\rm ln}(1-{r_{\rm g} \over
r})+{r_{\rm g} \over r}\cdot (1+{1\over 2}{r_{\rm g} \over r})\right)$$ 

and $f_1=f(r_{\rm NS})$ are the corrections due to general relativity (Muslimov
\& Tsygan).  In this context $r_{\rm g} = {{2\Gamma m_{ns}} \over c^2}$ denotes
the Schwarzschild radius of the neutron star. 

Assuming that the observed emission is indeed caused by coherent inverse Compton
scattering, we can calculate  the Lorentz factor of the emitting electron
current as a function of distance from the pulsar. 

\begin{equation}
\gamma(\omega _{\rm obs})=\left({{\epsilon _0 m_{\rm NS}} \over {n(r_{\rm em}
(\omega _{\rm obs}))} e^2}\right)^{1\over 3}\omega _{\rm obs}^{2\over 3}
\end{equation}

It varies from
13 (100 MHz) to 267 (30 GHz). 
Using the relativistically corrected values for distance, density and 
Lorentz factor we obtain an improved estimate of the dissipation scale 
$\lambda$ which is given by

\begin{equation}
\lambda={ L \over {P_{\rm coh}(r,\gamma)\cdot n(r) \cdot 2\pi r^2
\left(1-\cos(\phi)\right)}}
\end{equation}
Here $L$ and $\phi$ are the observed luminosity and opening angles of the pulsar
beam at a given frequency and $P_{\rm coh}$ means the emitted power per
coherently radiating particle. We estimated the luminosity as an upper limit
from the radio flux by assuming that the smaller cap area given by the
peak flux opening angle is filled with uniform emission
can be a good representation of the conal beam structure.
\begin{figure}[htbp]
 \centerline{\psfig{file=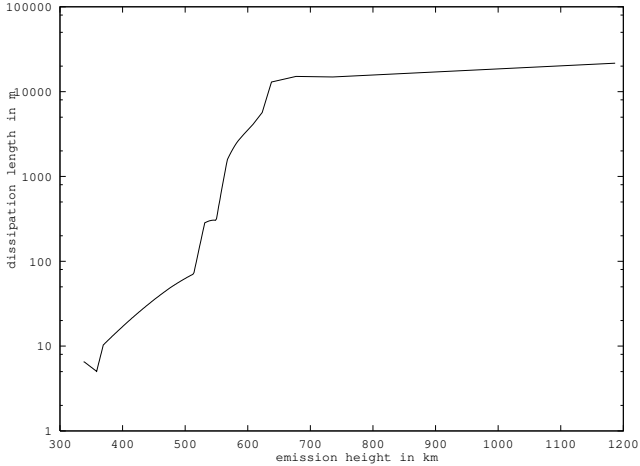,width=8.8cm,clip=}}
 \caption{\label{LAMBDA} Dissipation length $\lambda$ as a function of emission height.}
\end{figure}

Fig. 4 shows the development of the the dissipation
scale with emission height. After a steep rise at 500 km ($\nu>1000$~MHz), $\lambda$ stays
around 20 km up to a height 1100 km (100 MHz). Thus we have evidence for a quick onset of
a strong and stable coherence at a height of 600 km, lasting for another 600 km.
This is also borne out by the fact that the brightness temperatures of the source
vary only a little in this regime where most of the pulsar radio luminosity is created.
A typical fluctuation time scale of ${{\gamma\cdot\lambda/ c}= 1{\rm ms}}$
is associated with the dissipation scale below 1~GHz. 
Above that frequency it is expected to decrease monotonically to $20{\rm \mu s}$
at 20~GHz. These timescales are in good agreement with observational evidence
(Lange et al.~1998) where typical microstructure timescales were found to
range within $600-1200 \mu s$ for that pulsar at 4.85 GHz. There is also some
observational evidence for a decrease in microstructure timescales with
increasing frequency for the pulsars PSR B1133+16 and PSR B2016+28 (Lange et al.~1998). 

\section{Micropulse flux enhancement by beaming effects}

In this section we discuss anisotropy effects for pulse substructures which
may enhance the observed brightness temperatures by up to several orders of
magnitude. It is very unlikely that the flux enhancement only arises from a
growth of the dissipation region. For the nanosecond giant pulses of the Crab
pulsar this is even impossible (if the width of the radio emission region jumped
from only 0.001 $-$ which is extremely narrow $-$ to around 1 pulsar radius, the
peak would occur on a timescale of \mbox{$\tau_{\rm mic}\approx r_{\rm
NS}/c\approx 3\cdot 10^{-5}$ s}). But as such small time scales also mean highly
localized emission, the anisotropy of the elementary radiation process has to be
considered. For an isotropic process there is a simple connection between the
radiated power and the observed flux:

\begin{equation}
 S_{\rm obs}(\nu)=\frac{1}{4\pi d^2}\frac{\partial}{\partial \nu}P_{\rm
 rad}(\nu)
\end{equation}

where $P_{\rm rad}(\nu)$ is the total power emitted in frequencies below $\nu$,
whereas $d$ denotes the transverse extension of the radiation region. For a
rough quantitative estimate, we can set $P_{\rm rad}(\nu_{\rm crit})=L$, and
replace the derivative by a quotient. Thus we obtain the simpler equation

\begin{equation}
 S_{\rm obs}(\nu_{\rm crit})=\frac{L}{4\pi \nu_{\rm crit}d^2}\,.
\end{equation}

This flux is drastically enhanced by two effects, producing a strongly
anisotropic emission pattern. The first is the well-known relativistic
lighthouse effect (Rybicki \& Lightman, 1979). A relativistically moving source
of isotropic radiation emits most of the power in a narrow cone with an angular
opening of about $1/\gamma$ for $\gamma\gg 1$, as can easily be verified by
performing the Lorentz transformation. The exact result is

\begin{equation}
 \theta^\prime={\rm arctan}\left(\frac{1}{\gamma}\,
 \frac{\sin\theta}{\left(\cos\theta +\beta\right)}\right)
\end{equation}

with $\theta$ being the angle between the direction of the flux and source
velocity in the co-moving frame and $\theta^\prime$ denoting this angle in the
observer's system. For $\theta=90^{\rm o}$ and $\gamma\gg 1$ the approximation
$\theta^\prime\approx 1/\gamma$ is a good one.

\noindent Therefore we find a flux enhancement of

\begin{equation}
 \xi_1=\frac{4\pi}{\pi/\gamma^2}=4\gamma^2
\end{equation}
by the lighthouse effect.

Another anisotropy factor is the coherent radiation process itself, as has been
shown by Kunzl et al.~(1998b). To understand this effect, one has to
recall that coherent emission requires a phase coupling of the emitting
particles in one direction. For relativistic plasma emission, the preferred
direction obviously is the direction of the streaming velocity (forward
direction).

As long as the spatial dimension of the coherently radiating volume is small
compared to the emitted wavelength, the interference is still constructive even
under large angles to the forward direction. However, as soon as the extension
of the coherence region becomes comparable to the wavelength, the coherent
emitter can be seen as a phased  antenna field producing intrinsically beamed
emission.

The quantitative analysis of this effect has been performed by Kunzl {\em et
al.} (1998b). The first minimum appears under the angle

\begin{equation}
 \alpha_{\rm min}=\sqrt{1+\frac{2\pi}{T}}-1\,.
\end{equation}

Here $T:=2\pi d/\lambda$ with $d$ being the transverse extension of the emission
region. The angle of the minimum is much smaller than unity for $T\gg 1$.

Similar to the lighthouse effect, we can use this angular extension to estimate
the enhancement factor by coherence beaming. Thus the flux grows by a factor of

\begin{equation}
 \xi_2=\frac{4\pi}{\pi\alpha_{\rm min}^2}=
 \frac{4}{\left(\sqrt{1+\frac{2\pi}{T}}-1\right)^2}\,.
\end{equation}

A more realistic treatment of the geometry uses a cylindrically symmetric
soliton with a density profile proportional to ${\rm cosh}^{-2}(R_0/R)$ where
$R_0\approx \lambda_{\rm D}$ is the typical length scale. Numerical calculations
for this structure show a much stronger anisotropy (Kunzl et al.~1998b)

Combining the two anisotropy effects we find the total beaming factor

\begin{equation}
 \label{beamges}
 \xi:=\xi_1\xi_2=16\gamma^2\left(\sqrt{1+\frac{2\pi}{T}}-1\right)^{-2}
\end{equation}

which, even with quite moderate parameters (like $\gamma=10, x_{\rm em}=50,
N=10^{13}$ for a typical pulsar), causes an enhancement of some $10^3$.
Therefore it can easily explain strong substructures, and micropulse brightness
temperatures that are some $10^3$ times above the mean value (Boriakoff, 1992).

As these beaming effects apply only to the elementary emission process, but do
not enhance the mean flux significantly, it is an
important check on whether the model can reproduce the observed brightness
temperatures of up to $10^{31}$ K (Hankins, 1996) observed in {\em
giant pulses} of the Crab pulsar.

Therefore we take the maximum possible number of coherently radiating particles,
and the corresponding beaming effects. Let $A$ be an arbitrary cross
section. Then starting with eq.~(\ref{Helltemp}) one can express the integrated
flux $I:=I_\nu\Delta \nu=L_{\rm {radio, A}}/A$ by the brightness temperature:

\begin{equation}
 I:=\frac{L_{\rm {radio, A}}}{A}=\frac{2\pi\nu^3k_{\rm B}T_{\rm B}}{c^2}\,.
\end{equation}

Inserting the critical frequency of the Crab pulsar $\nu=\nu_{\rm crit}=160$ MHz
and $T_{\rm B}=10^{31}$ K, the integrated flux is

\begin{equation}
 I=3.95\cdot 10^{16}\,\frac{\rm W}{\rm m^2}\,.
\end{equation}

As this value means the beamed flux, the actual power per area is only

\begin{equation}
 \label{Ireal}
 I_{\rm real}:=\frac{I}{\xi}=2.47\cdot 10^{15}\frac{\rm W}{\rm m^2}\,\gamma^{-2}
 \left(\sqrt{1+\frac{2\pi}{T}}-1\right)^2\,.
\end{equation}

Now we can compute the minimum thickness of a layer producing the power per
cross-section. To obtain the lower limit we take the strongest dissipation field
possible (see eq.~(\ref{Ediss})) and assume a coherence cell with a lateral
extension of $c/\nu_{\rm p}$, so that $T=2\pi\gamma$.

The total dissipated power from a volume with the cross-section $A$ and the
radial extension $d$ is

\begin{equation}
 \label{Pdiss}
 P_{\rm diss}\stackrel{!}{=}I_{\rm real}A = E_{\rm diss}\,d\,j\,A\,.
\end{equation}

For a relativistic Goldreich-Julian current $j=j_{\rm GJ}=n_{\rm GJ}(x)ec$
eq.~(\ref{Pdiss}) can be solved for $d$. Inserting the Crab pulsar parameters
($P=33.4$ ms, $B=3.8\cdot 10^8$ T, $\nu=160$ MHz, $x_{\rm em}=80$) and
additionally using eq.~(\ref{gamomega}) we obtain

\begin{equation}
 d=3.85\,{\rm m}\left(\frac{T_{\rm B}}{10^{31}\,{\rm K}}\right)
\end{equation}

which corresponds to a time scale of

\begin{equation}
 \tau:=\frac{d}{c}=1.28\cdot 10^{-8}\,{\rm s}\left(\frac{T_{\rm B}}
 {10^{31}\,{\rm K}}\right) = 12.8\,{\rm ns}\left(\frac{T_{\rm B}}{10^{31}
 \,{\rm K}}\right)\,.
\end{equation}

Thus we can expect to see giant pulses on nanosecond timescales, with brightness
temperatures above \mbox{$10^{30}$ K}, from far inside the light cylinder in the
Crab pulsar, although the relative emission height will be considerably larger
than for average pulsars.

\section{Densities in the radiation region}

Here we briefly discuss if the model described above could also work for
particle densities significantly above or below the GJ value. We will show that
even if there were a mechanism producing a current density and particle energies
matching the requirements to produce the low radio frequencies, such a scenario
would be very unlikely.

Assuming that the particle density was much larger than the value predicted by
Goldreich \& Julian (1969), the Lorentz factor corresponding to a certain
frequency would drop according to (\ref{gamomega}). Especially for young pulsars
this would quickly require $\gamma < 1$ which is, of course, not possible (even
$\beta\gamma\geq 1$ should be fulfilled as the particles are definitely
relativistic). This important fact has already been pointed out by Kunzl {\em et
al.} (1998a).

\smallskip

A relativistic current significantly below the GJ current (meaning that the
particle density is smaller than $n_{\rm GJ}$ faces even more difficulties. On
the one hand, accelerating fields are incompletely shielded (the equilibrium
Lorentz factor where acceleration and radiation losses $-$ now including coherent
radio emission $-$ balance is extremely high, so coherence is highly doubtful).
Furthermore the one-dimensional approach would no longer account as a permanent
strong deviation from ideality allows efficient $\vec{E}\times\vec{B}$ drift.
But also the existence of strong $\gamma$ -pulsars provides an argument against
this solution:

As we observe {\em pulsed} $\gamma$-emission of pulsars whose high energetic
luminosity reaches up to 10\% of the spindown luminosity, some mechanism must
efficiently convert {\em particle energy} into (hard) radiation (Kanbach,
2001). But the
potential difference a particle can suffer is limited by the polar cap
potential. Since the power dissipated by a relativistic GJ current flowing over
the entire polar cap area with the polar cap voltage is comparable to the
spindown power of the neutron star, a large fraction (well above 10\%) of the
generator's power must be pumped into particles. This, of course, is only
possible for currents of about the GJ current and thus for particle densities
comparable to the Goldreich-Julian value.

\section{Discussion}

We have presented the outline of a model which may explain the creation of radio emission
within the energy constraints of the observationally deduced
emission heights of 50-100 pulsar radii. (e.g. Blaskiewicz et al.~1991; 
Taylor et al.~1993; Kijak and Gil 1997,1998;  Kramer et al.~1997). 
Several other pulsar models fail to explain this feature,
e.g.~the model of Lyutikov et al.~(1999) which uses Cherenkov drift
resonance, but predicts the radiation zone to be in some 1000 pulsar radii.

One of the main differences to the standard model is that we do not assume
an inner gap which accelerates particles to ultrarelativistic energies and
is saturated by a pair avalanche. Instead we suggest a conducting wire model with a
quasi- stationary, low relativistic current that suffers fluctuations on short
time scales. Such a model is supported by calculations on particle densities
available from the polar cap by thermal and field emission (Jessner et al.~2001)
which show that inner gap models have severe difficulties in producing particles
energetic enough to start pair cascades for parallel or nearly parallel
rotators. 

We showed that the brightness temperatures for the mean pulse can be reproduced
by our model and also micropulse fluxes can be explained by appropriate
beaming assumptions.

The radiation process in the framework of such a model could be some
relativistic plasma emission like the ALAE process described by Melrose
(1978). This mechanism requires mildly relativistic particles to be efficient.
 As shown also in Kunzl et al.~(1998a) any plasma process responsible for 
the low frequency emission in the Crab pulsar must work with mildly 
relativistic particles. Von Hoensbroech et al.~(1998) showed that in a Goldreich-Julian
magnetosphere with low relativistic particles propagation effects can also
qualitatively explain the observed polarization properties.

Finally we would like to note that our model does not rule out pair production
in general. As can be seen from several observations of the Crab pulsar a dense
particle wind is needed which transports energy to the nebula. The calculations
of Hirotani and Shibata (1999) show that a significant fraction of the vacuum
potential drop can occur in an outer gap if the X-ray photon density is not too
high. For a detailled geometric examination of the outer gap structure see Cheng et al.~(2000).

It has been shown that for reasonable parameters an outer gap model could easily
explain the observed energy and particle fluxes in the pulsar wind. Recently
Crusius et al.~(2000) have shown that an outer gap model can explain the
observed spectrum and energy output of the Crab pulsar from the infrared to
X-ray frequencies which provides another hint that the main dissipation region
is in the outer magnetosphere. 

We gratefully acknowledge constructive discussions with J. Arons, D.B. Melrose 
M. Kramer, D. Mitra and A. von Hoensbroech. In addition D. Mitra and M. Kramer provided us with access to 
their radio flux- and emission height measurements for PSR 0329+54 for 
which we are particularly grateful.


\clearpage

\end{document}